\documentclass[aps,prc,twocolumn,floatfix,superscriptaddress]{revtex4}
 
\usepackage{epsfig}
\usepackage{amsmath}
\usepackage{color}
 
\usepackage{graphicx}
\begin{document}
 
\title{Gap states controlled transmission through 1D Metal-Nanotube junction}

\author{D. Talukdar}
\email{dtalukdar@ntu.edu.sg}
\affiliation{Saha Institute of Nuclear Physics, I/AF Bidhannagar, Kolkata 700 064, India }
\affiliation{Nanoscience Center, Department of Physics, P.O.Box 35, FI-40014 University of Jyv\"askyl\"a, Finland}
 
\author{ P. Yotprayoonsak}
\affiliation{Nanoscience Center, Department of Physics, P.O.Box 35, FI-40014 University of Jyv\"askyl\"a, Finland}
\author{C.D. Mukherjee}
\affiliation{Saha Institute of Nuclear Physics, I/AF Bidhannagar, Kolkata 700 064, India }
\author{K. K. Bardhan}
\affiliation{Kalpana Chawla Centre for Space and Nano Sciences, 3F Swamiji Nagar, Kolkata 700 030, India}
\author{B. Karmakar}
\affiliation{Saha Institute of Nuclear Physics, I/AF Bidhannagar, Kolkata 700 064, India }
\date{\today}
 
 
\maketitle
 \textbf{Understanding the nature of metal/1D-semiconductor contacts such as metal/carbon nanotubes is a fundamental scientific and technological challenge for realizing  high performance transistors\cite{Francois,Franklin}. A Schottky Barrier(SB) is usually formed at the interface of the $2D$ metal electrode with the $1D$ semiconducting carbon nanotube.  As yet, experimental\cite{Appenzeller,Chen, Heinze, Derycke} and numerical \cite{Leonard, Jimenez} studies have generally failed\cite{Svensson} to come up with any functional relationship among the relevant variables affecting carrier transport across the SB owing to their unique geometries and complicated electrostatics. Here, we show that localized states called the metal induced gap states (MIGS)\cite{Tersoff,Leonard} already present in the barrier determines the transistor drain characteristics. These states  seem to have little or no influence near the ON-state of the transistor but starts to affect the drain characteristics strongly as the OFF-state is approached. The role of MIGS is characterized by tracking the dynamics of the onset bias, $V_o$ of non-linear conduction in the drain characteristics with gate voltage $V_g$.  We find that $V_o$ varies with the zero-bias conductance $G_o(V_g)$ for a gate bias $V_g$  as a power-law: $V_o$ $\sim $ ${G_o(V_g)}^x$ with an exponent $x$. The origin of this power-law relationship is tentatively suggested as a result of power-law variation of effective barrier height with $V_g$, corroborated by previous theoretical and experimental results\cite{Appenzeller}. The influence of MIGS states on transport is further  verified independently by temperature dependent measurements. The unexpected scaling behavior seem to be very generic for metal/CNT contact providing an experimental forecast for designing state of the art CNT devices.}

            In nanoscale FETs depending upon the band bending at the metal-nanotube interface a Schottky Barrier(SB) is formed at the source and drain contacts. The charge injection in such a device is mainly determined by tunneling across the SB, whose width can be modulated by the gate voltage, $V_{g}$. The resulting current through the SB (a function of $V_{g}$) is usually given by the Landauer-B$\ddot{u}$ttiker formula\cite{Datta}:
\begin{equation}
I(V)=4e/h \int P(E,V)[F(E-eV)-F(E)] dE
\label{eq:LB}
\end{equation}
where $P(E,V)$ is the transmission probability across the junction at bias $V$ for carriers having energy $E$. Linear $I_{ds}$-$V_{ds}$
output characteristics are expected for this formula at low drain bias for ballistic tubes or wires with no defects. For a fixed gate bias $V_g$, however one expects deviations from linear behavior with increasing drain bias $V_{ds}$  at some onset bias, $V_o$ due to lowering of barrier heights leading to increased tunneling currents. At the onset of non-linear conduction $V_o$, the electronic energy $eV$ will be almost equal to the effective  barrier height $ \phi_{SB}$  resulting in increased tunneling probability\cite{Simmons}:
\begin{equation}
 \Phi_{SB} (V_g) \sim  \left|e V_{ds}\right|_{V_o}         
\label{eq:ON}
\end{equation}
where $[V_{ds}]_{V_o} $ is the source drain onset voltage and $\phi_{SB}(V_g)$ is the effective SB height which is a function of the $V_{g}$. Eq. \ref{eq:ON} implies that there is a voltage scale for non-linearity in the system and for each $V_g$ there will be a corresponding onset drain bias, $V_o$ which can be used as a measure of the effective barrier height $\phi_{SB}$."Effective barrier height" $\phi_{SB}$ is assumed as the flatband conditions are not applicable at finite temperatures, gate and drain bias and also contains the effect of other details e.g., metal-nanotube coupling as a function of curvature.
In literature the standard approach \cite{Chen,Appenzeller} so far has been to calculate the effective SB height by self-consistently solving the Poisson and Schr$\ddot{o}$dinger equation and experimentally by studying the device ON-current, $I_{on}$. As non-ohmic behavior is usually obtained when there is increased tunneling due to barrier modification by the source drain bias $V_{sd}$, it is quite evident that the dynamics of the barrier change is contained in the drain characteristics $I_{sd}$-$V_{sd}$. The threshold bias for non-linearity $V_o$ can thus be used to characterize the SB. The question then is: Is there any general functional relationship between the relevant parameters affecting this change which should in principle be valid for all 1-D nanotransitors irrespective of device geometry?
  
             The question can be addressed by analyzing the non-linear conduction properties, in particular the onset bias, $V_o$ as a function of gate voltage, $V_g$. To find the  onset or the crossover bias, $V_o$ there are mainly two methods used in literature. In the first method a crossover point $V_o$ is chosen at which the conductance deviates significantly from the zero field conductance $G_o$ by some arbitrary fraction $\epsilon$, such that the conductance $G=G_o (1+\epsilon)$. In the second method\cite{Talukdar11,Talukdar12} based on the existence of a single voltage scale $V_o$, a set of $I-V$ curves can be characterized by a number $x$, called the nonlinearity exponent. In this scaling approach, the conductance $G (M,V)$ is given by the scaling relation:
\begin{equation}
{G(M,V) \over G(M,0)} = Q \left ({V \over V_o} \right ),
\label{eq:scaling}
\end{equation}
where  \textit{Q} is a scaling function and the bias scale $V_o(M)$ at each variable value, \textit{M} (gate voltage) is given by the phenomenological relation
\begin{equation}
V_o (V_g) = A \; {G_o}^{x},
\label{eq:fscale}
\end{equation}
where $G_o(M) = G(M,0)$ is the conductance at zero bias at gate voltage $V_g$ and $A$ is a constant. After evaluating $V_o$ from the first method  the scaling can be applied for the data following Eq. \ref{eq:scaling}. Thus, both the methods are  essentially equivalent \cite{Talukdar11,Talukdar12}  for determination of the nonlinearity exponent $x$. For CNTFETs the barrier is thin near the ON-state gradually becoming thicker towards the OFF-state. Accordingly, the threshold bias for non-linearity $V_o$ in the $I_{sd}-V_{sd}$ characteristics is also expected to vary systematically with $V_g$. In the OFF state the onset bias for non-linearity should start at a higher voltage, $V_o$ than near ON-state where the corresponding  $V_o$ should start earlier. Consequently, from Eq. \ref{eq:fscale} we expect that the non-linearity exponent to be negative for CNTFETs.

\begin{figure}[htbp]
\includegraphics[width=9cm]{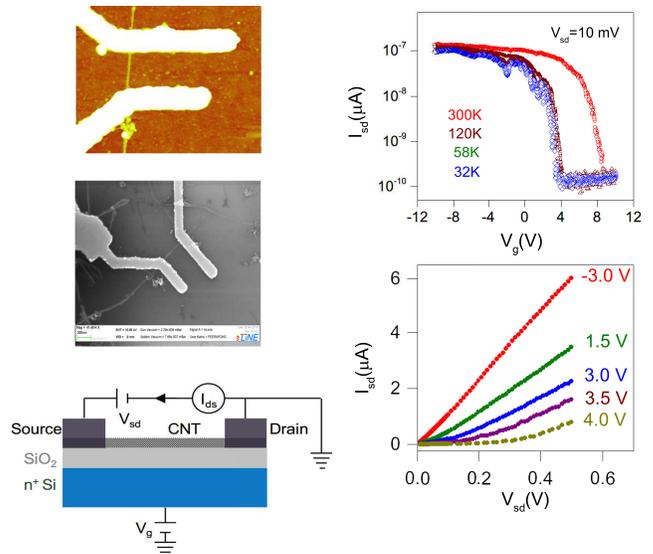}
\caption{ \textbf{Figure 1. The transfer and ouput characteristics of the CNT transistor} (a)Schematic and AFM image of the bottom gated CNT transistor with 300 nm oxide thickness. (b)Transfer characteristics of a CNTFET for different temperatures taken at $V_{ds}$=10 mV. $I_{on}$ changes weakly as temperature is lowered signifying increased tunneling at lower temperatures. (c) Output $I_{sd}$-$V_{sd}$ characteristics for the same device at 120 K at different values of gate voltage in the subthreshold region. The curves become increasingly non-linear while approaching the OFF-state.}  
\label{fig.1}
\end{figure}			 			        
 
                A typical CNT device architecture with a representative AFM and SEM image is shown in Fig. 1(a)-(c). The CNT is lying on a 300nm $SiO_2$ layer with  Si as backgate connected with Pd contacts (see Methods for details). All measurements were done on small diameter tubes with d$\le$1.5nm as the SB height is inversely proportional to the nanotube diameter\cite{Chen}. More than four devices were studied at different temperatures in this study. Note that usually the fabrication process incorporates some additional tunneling barrier in the devices in addition to the SB already present\cite{Svensson}. The thickness of the material and diameter of   the CNT is a crucial parameter for the SB of the CNTFET as they influence the electric field at the contact which controls its switching action.

              Figure 1 (d) presents the variation of $I_{sd}$ with $V_g$ as a function of temperature in a backgated CNT transistor with a channel length of 300 nm and diameter 1.4 nm. Resistance more than 6.5 k$\Omega$ (quantum limit h/4$e^2$$\approx$4.5k$\Omega$) or low value of ON-current are indicative of the existence of SB at the contacts for p-type hole transport\cite{Javey}. Note that tunneling currents $I_{sd}$ increases as $V_{g}$ becomes more negative  indicating the thinning of barrier for current injection switching the transistor from OFF to ON state. Here, the current  modulation by $V_{gs}$  directly translates into corresponding change in $\phi_{SB}$. Below 120 K the $ON$-current becomes independent of temperature indicating that the dominant contribution to the current injection is due to tunneling with negligible contribution from thermal assisted tunneling\cite{Appenzeller, Martel, Svensson}. At this temperature $I_{sd}$-$V_{sd}$ characteristics  in Fig. 1 (e) contains complete information about the transport mechanism through the SB with minimal interference from thermal excitations.  In the ON-state when the SB is  thin the current increases rapidly with $V_{sd}$ and is linear while near the OFF-state the rate of current increase is relatively slower and non-linear with $V_{ds}$. Although, the $I_{sd}$-$V_{sd}$ characteristics usually give a qualitative understanding of the situation, for developing a framework to analyze the non-linear transport data, the equivalent conductance $G$-$V_{sd}$ description is usually more useful as described below.

 \begin{figure}[htbp]
\includegraphics[width=9cm]{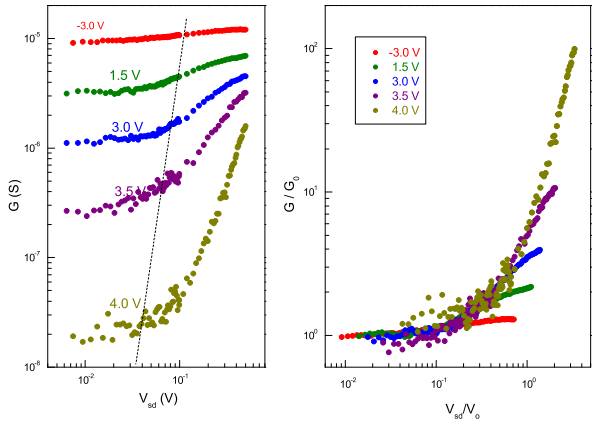}
\caption{\textbf{Figure 2. Equivalent characterization of CNT transitors} (a) Variation of the conductance vs. drain bias of the same data of figure 1 (c) at different values of gate bias. The dotted line schematically shows the motion of onset field $V_o$ as the gate bias $V_{gs}$ is changed.  The saturation of conductance seems to have a threshold of $\sim$ 160 mV which is due to emission of optical phonons.  (b) The scaling of the same data as in figure 2a to achieve data collapse. There seems to be an envelope for the scaling curve from where deviations are due to optical phonon scattering.}
\label{fig.1}
\end{figure}

              The systematics of the evolution of $I_{sd}$-$V_{sd}$ drain characteristics with changing $V_g$ is better illustrated by replotting them as $G$-$V_{sd}$ shown in Fig. 2 (a). The $G$-$V_{sd}$  curves are qualitatively identical, initially linear and after a threshold onset voltage $V_o$ start becoming non-linear. This general behavior of conductance for all measured tubes can be understood in the following manner.  With increasing  barrier thickness i.e., increasingly positive $V_g$ the $G_o(V_g)$-$V_{sd}$ curves tend to become non-ohmic at smaller  drain bias, $V_{sd}$. The systematics can be put in perspective by tracking the onset-bias for non-linear conduction $V_o$ as a function of $V_g$, shown by the dotted line in the figure. It is clear that onset voltage $V_o$ is intricately related to the zero-bias conductance $G_o(V_g)$ modulated by the gate bias. However the direction of movement of  $V_o$  with $G_o(V_g)$ is opposite to what is expected from Eq. (4). A distinctive feature of the $I_{sd}$-$V_{sd}$ characteristics is the low values of onset voltages $V_o$, a measure of the effective barrier heights, varying between $20-100$ mV. This is much lower than the effective barrier values obtained usually in literature which varies between $100-400$ $meV$\cite{Appenzeller,Chen, Heinze}. The cause of this discrepancy will be discussed in a later section.

               It is clear from Figure 2(a) that the $G(V_{sd})$ curves for different $V_g$ can be collapsed into a single curve  using the scaling procedure discussed earlier. The results of this scaling are illustrated in Fig. 2(b) where  $G/G_o$ is represented as a function of $V/V_o$. The detailed scaling procedure is described in greater detail elsewhere\cite{Talukdar11,Talukdar12}. The remarkable data collapse underlines the fact that there is a field scale for non-linearity in CNT transistors. The existence of the scaling is very significant here as it implies that the $I_{sd}-V_{sd}$ curves for a CNT transistors are identical differing by only a constant. The deviations from the scaled curve at high-bias ($eV_{sd}$$\ge$0.16 eV) is due to the well known electron backscattering  due to emission of zone-boundary or optical phonons \cite{Yao, Back, Dresselhaus} in single walled carbon nanotubes. The scaling holds for all the studied tubes and different temperatures.

 \begin{figure}[htbp]
\includegraphics[width=9cm]{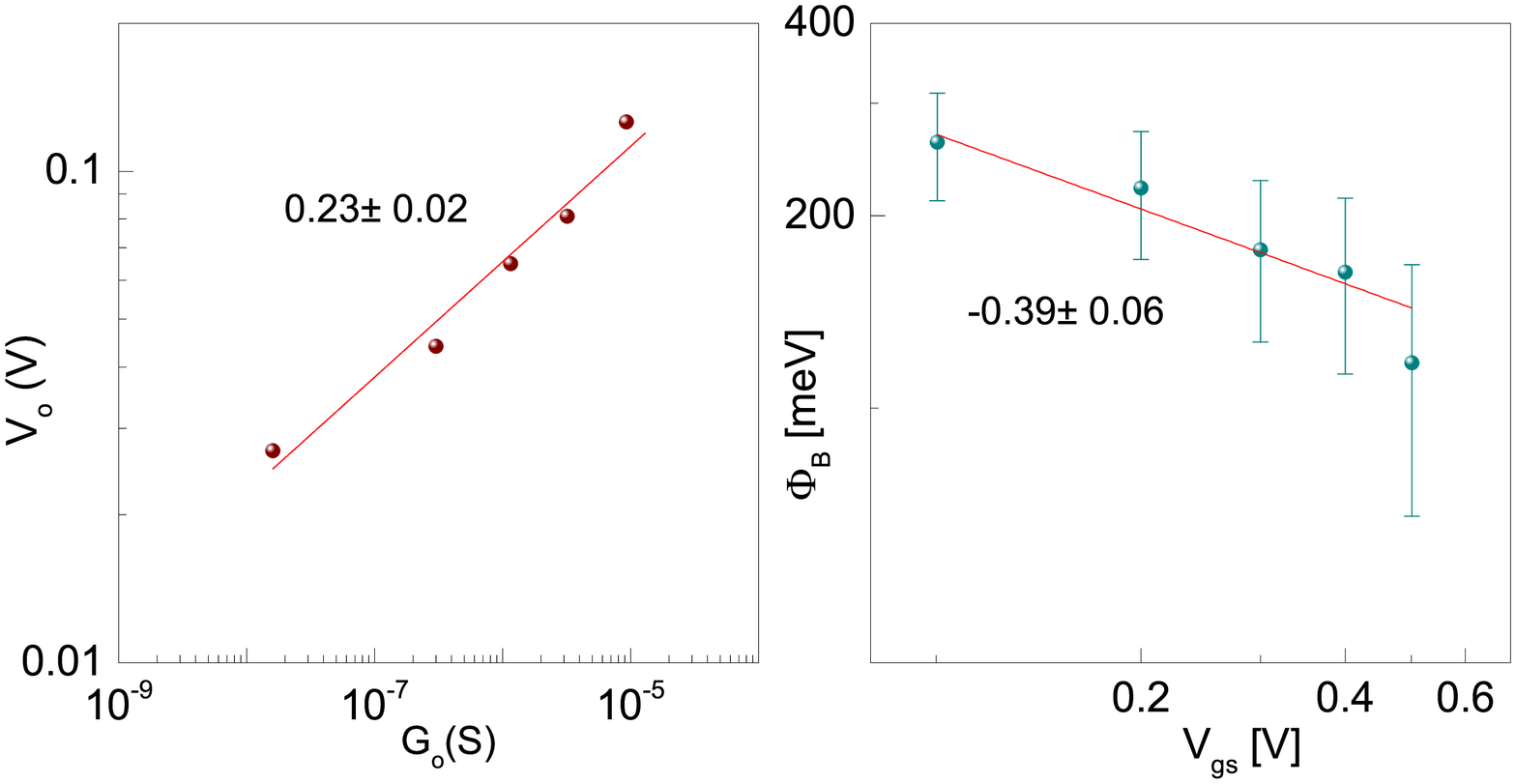}
\caption{ \textbf{Figure 3. Determination of functional relation ships for characterization of the Schottky barriers for hole injection:} (a) Plot of onset voltage $V_{o}$ with zero bias conductance $G_{o}$ which is a function of gate voltage in the subthreshold region. The linear fits represent a power law between the two with the slopes mentioned.  (b) log-log plot of effective schottky barrier height $\Phi_B$ vs. $V_{gs}$ representing  competition between thermoionic emission and quantum mechanical tunneling obtained from Ref. \cite{Appenzeller}. The plot indicates that the barrier height changes as a power law with gate voltage in CNTs.}
\label{fig.1}
\end{figure}

            To examine the scaling behavior, we plot $V_o$ and $G_o$ on a log-log plot as shown in Fig. 2(a). The solid line indicates that there exists a power-law relationship between $V_o$ and $G_o(V_g)$ i.e., 
\begin{equation}
V_o \sim {G_o(V_g)}^{x}
\label{eq:}
\end{equation}
with a positive non-linearity exponent $x$ of $0.23\pm0.02$ as suggested by the orientation of the dotted line. The existence of the power-law automatically implies at high fields the conductance goes as a power-law with bias (see appendix), which is however masked in this case as conductance starts to saturate much earlier due to optical phonon scattering (0.16eV)\cite{Yao, Back}. We extract the values of  non-linearity exponent $x$ for this tube at different temperatures (supplementary information). Surprisingly, the sign of the exponent is positive for all the measured tubes at different temperatures.  Positive exponent implies that for thinner barriers (ON) higher drain voltage is required for the onset of non-linear conduction than near the OFF-state (thicker barriers). Existence of positive exponents in these tubes is counterintuitive to the expected negative exponent as discussed earlier in Eq. (4). With regard to this, it is of prime importance to know the dynamics of SB change with $V_g$, as $V_o$ depends on the nature of SB which determines $G_o$, implicitly dependent on $V_g$. To understand this dynamics, we note that Appenzeller et. al.\cite{Appenzeller}  simulated and experimentally determined the effective barrier heights as a function of $V_g$ using the non-equilibrium Green's function (NEGF) formalism and a thermal activation model respectively. Strikingly, their data shown in Fig. 3(b) indicates that the effective barrier height $\Phi_B$ of the metal-CNT interface  also varies as a power-law with $V_{g}$, implying the power-law behavior is  strongly linked to the modulation of the barrier.

 \begin{figure*}[htbp]
\includegraphics[width=13cm]{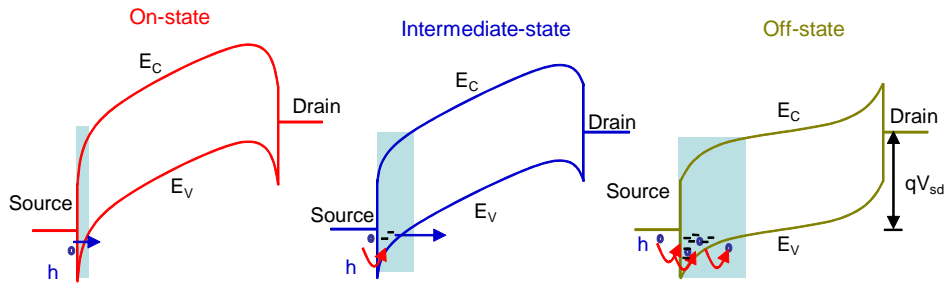}
\caption{\textbf{Figure 4. Schematic for band alignment for different gate voltages at metal/nanotube interface }(a) ON-state: Qualitative response of conduction through the schottky barrier at the metal/nanotube interface at a gate voltage close to threshold voltage where the transistor ON. The barrier  shown by the colored band is sufficiently thin assisting direct tunneling (b) Intermediate state: The band bends for a gate voltage below the threshold where the SB is sufficiently thick making the MIGS slowly relevant for transport (c) OFF state: The SB is the thickest and the transport through the localized MIGS states dominates.}
\label{fig.4}
\end{figure*}

            To understand this anomalous behavior and account for the origin of functional relationships obtained we need to examine the different band bending scenarios in the CNT device for different gate bias conditions ranging from OFF to ON. In Fig. 4(a) the CNT-device is in an ON-state in 4(b) at an intermediate gate bias and for  4 (c) the device is OFF. As hole transport is considered we mainly concentrate on the the valence band. The shaded bands represent the SB on the source and drain sides respectively. In figure 4 (a) when the transistor is ON the barrier thickness is reduced leading to direct tunneling and subsequently lower values of $V_{ds}$ for non-linear conduction. For close to OFF-state the barrier is considerably thicker resulting in lower tunneling currents in turn requiring larger $V_{ds}$ for non-linear conduction. However, the experimental situation is completely opposite. The discrepancy in the non-linear behavior, manifested distinctly in the dependence of  $V_o$ on the barrier thickness, and thus conductance $G_o$ demonstrates the role of the localized states for transport. These states located in the metal-CNT interface are the  Metal Induced gap states (MIGS)\cite{Tersoff,Tersoff1,Heine,Louie} whose role in carrier transport was not recognized in literature. These states arise as a result of electronic wave functions being abruptly terminated at the metal-CNT junction and extends only up to a few nanometer into the CNTs getting localized close to the interface. These interface states have little or no effect on electrical behavior near the ON-state when the barrier is thin as transport is through direct tunneling through the barrier leading to a particular onset bias for non-linearity $V_o$. However, near the OFF-state when the barrier is sufficiently thick the dominant conduction mechanism crosses over from direct tunneling to transport through localized states in the barrier as function of barrier thickness. This can explain the lower thresholds for non-linear conduction with increasing barrier heights as shown earlier which is a characteristic of tunneling through localized states\cite{Xu}. The localized states at the interface states thus play a crucial role for controlling the device characteristics in CNTFETs and sets the scale for non-linearity in these devices. The MIGS states present in the barrier can also explain the asymmetry of the $I_{ds}-V_{ds}$ characteristics commonly obtained for carbon nanotube transistors \cite{Li,Khoo} which is due to different configurations of MIGS states present in each metal-semiconductor barrier on the source and the drain side. 
            
            The origin of the power-law relationship between $V_o$ and $G_o$ might arise due to the variation of barrier height $\Phi_B$ as a power-law with $V_{g}$ having an exponent 'y' as obtained by  Appenzeller et. al. Physically, the exponent 'y' indicates a ratio of $V_{gs}$ and the electric field at the contact. As $G_o$ depends on the the configuration of localized states at the interface varying with $V_{gs}$ as a power-law it indicates the ratio of $V_{g}$ and  the distribution of localized states $D(E)$ at the interface.

\begin{figure}[htbp]
\includegraphics[width=4.5cm]{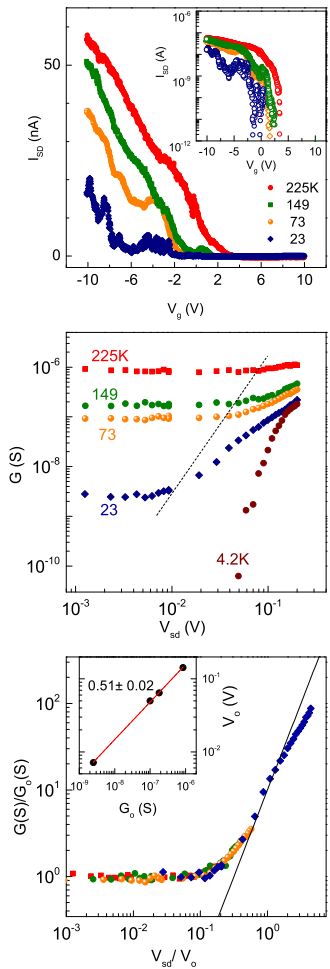}
\caption{\textbf{Figure 5. Temperature dependence for a CNTFET} (a) The experimental transfer chracteristics  for different temperatures at $V_{ds}$= 10 mV. $I_{on}$ varies by an order of magnitude with temperature. The inset shows the same transfer characteristics with $I_{ds}$ values in logarithmic scale. (b) Output characteristics plotted as $G-V_{sd}$ at different temperatures. The onset voltage $V_o$ start decreasing with lowering temperatures shown by the dotted line. At 4.2K the $G-V_{sd}$ is completely non-linear. (c) Scaling of the same data leading to data collapse as shown. $G_o(T)$ and $V_o(T)$  are the zero bias conductance and onset bias at some temperature T. The inset shows the log-log plots of  $G_o(T)$ and $V_o(T)$ with linear fits and slopes mentioned. }
\label{fig.5}
\end{figure}

             To verify the physical reasoning described above we made additional temperature dependent measurements on our devices. In Figure 5 (a) the gate characteristics of a device in both linear- and log-scale is shown. The drain characteristics $I_{sd}$-$V_{sd}$ at zero gate bias is shown in Figure 5 (b). Zero gate bias is chosen so as to keep the barrier without being affected by the gate which will enable us to see the intrinsic barrier characteristics. Also for illustration of this method it is advantageous to take the drain characteristics $I_{sd}$-$V_{sd}$ at zero gate bias as they are more non-linear. It can be seen from the figure that the drain curves start to become progressively non-linear with decreasing temperature. To find the onset bias $V_o$ shown by the dashed line, we can use the same scaling procedure described above  obtain a perfect data collapse for normalized conductance $G(T,V)/ G(T,0)$ vs. the normalized bias $V/V_o$. The bias scale $V_o$ appears to be a power-law function of the Ohmic conductance and decreases with decreasing temperature as shown in the inset of Fig. 5c (circles). This is similar to the situation where the conductance was modulated by $V_{g}$. This can be understood qualitatively in the following manner. As the temperature decreases the  contribution of thermally assisted tunneling for transport is reduced. Initially $I_{sd}$-$V_{sd}$ curves are therefore  ohmic at high temperature as the main contribution to carrier injection is by thermal activation. However, with decreasing temperature, thermal activation is reduced significantly and the effect of localized states in the barrier can no more be neglected. With progressive decrease of temperature more and more localized states in the barrier (MIGS) start taking part in conduction leading to lower onset bias $V_o$ with decrease of temperature. However, it is still not clear why  $V_o$  varies with conductance $G_o$ as a power law. It must be mentioned here that the curve at $4.2K$ is not used for the scaling as the conduction mechanism is different in this case due to the onset of coulomb blockade\cite{Postma,Park} effects. 
  
                  In conclusion, our experiments provide first clear experimental evidence that even though the transistor action observed in nanoscale CNTFETs is based on transport across a SB barrier the $I_{sd}$-$V_{sd}$ curves can only be explained if we consider the barriers to contain localized MIGS states. Using experiments and comparing with previous simulations and experimental data from literature we suggest that the effective SB heights and threshold bias for non-linearity in CNTFETs are modulated by the gate voltage/source-drain voltage in a power-law fashion  with a nonlinearity exponent $x$. Further, the formalism can explain the discrepancy commonly observed between theoretical and experimental SB values for any low dimensional material. Our method of analyzing nonlinear transport data can also be used to characterize the cut off working linear regime voltage  of a CNTFET or nanowire FET device.

\textbf{Methods}
 In our studies the SWNT devices used here fabricated by spin depositing suspension of nanotubes with (1,2)-dichloroethane onto highly p-doped Silicon wafers covered with 300 nm thick thermally grown $SiO_2$. AFM (atomic force microscope) was used for locating the tubes and standard e-beam lithography steps were used to electrically contact the nanotubes. Pd of thickness 25 nm was deposited over the CNTs in order to improve the contact resistance. Low temperature transport measurements were performed in a RF-shielded room using home made dipstick.

\textbf{Appendix} 
The field-dependent conductance as shown in Figure. 2a and 5b clearly show that at high fields conductance becomes independent of the variational parameter (Gate bias and Temperature respectively). If the conductance is continued to be described by Eq. (\ref{eq:scaling}) it is obvious that at large fields the conductance varies with bias as a power-law: $G$$\sim$$F^z$ where '$z$' is the high field exponent and $x=1/z$.

\begin{acknowledgments} 
We acknowledge helpful discussions with Prof. M. Ahlskog. 
\end{acknowledgments}

\end{document}